\documentclass[galaxies,article,submit,moreauthors,10pt,letterpaper]{mdpi_no_lineno}

\pdfoutput=1

\firstpage{1} 
\articlenumber{x}
\doinum{10.3390/------}
\pubvolume{xx}
\pubyear{2016}
\copyrightyear{2016}
\externaleditor{Academic Editors: Jose L. G\'{o}mez, Alan P. Marscher and Svetlana G. Jorstad}
\history{Received: date; Accepted: date; Published: date}

\Title{PARSEC-SCALE STRUCTURE AND KINEMATICS OF FAINT TEV HBLS}

\Author{B. Glenn Piner $^{1}$ and Philip G. Edwards $^{2,*}$}
\AuthorNames{Glenn Piner and Philip Edwards}

\address{%
$^{1}$ \quad Whittier College; gpiner@whittier.edu\\
$^{2}$ \quad CSIRO Astronomy and Space Science; philip.edwards@csiro.au}

\corres{Correspondence: philip.edwards@csiro.au; Tel.: +61-2-9372-4717}

\abstract{
We present new multi-epoch Very Long Baseline Array (VLBA) observations of a set of 
TeV blazars drawn from our VLBA program to monitor all TeV-detected high-frequency 
peaked BL Lac objects (HBLs) at parsec scales. Most of these sources are faint in the 
radio, so they have not been well observed 
with VLBI by other surveys. Our previous measurements of apparent jet speeds in
TeV HBLs showed apparent jet speeds that were subluminal or barely 
superluminal, suggesting jets with velocity structures at the parsec-scale. Here we 
present apparent jet speed measurements for eight new TeV HBLs, which for the first 
time show a superluminal tail to the apparent speed distribution for the TeV HBLs.
}


\begin{document}

\section{Introduction}

At TeV energies (10$^{12}$eV), three orders of magnitude higher than
those studied by satellite-based detectors, gamma-ray astronomy is
conducted with ground based telescopes such as H.E.S.S., VERITAS and MAGIC.
Over 175 TeV gamma-ray
sources have now been catalogued (http://tevcat.uchicago.edu/), with
over one third of these being extragalactic objects. The majority of
these (46 of 69) are classified as HBL (High-frequency–peaked BL Lac)
objects, for which the synchrotron peak of the Spectral Energy
Distribution (SED) lies at frequencies above 10$^{16.5}$\,Hz.

Many of the well-studied HBL TeV
sources have shown dramatic variability in their gamma-ray emission
\cite{kra04,aha07a}. The most rapid variations suggest extremely small
emitting volumes and/or time compression by large relativistic Doppler
factors of up to $\sim$100
and challenge our understanding of relativistic jets \cite{aha07a,beg08,ghi08}.

The only way to directly obtain information on the parsec-scale
structure of these blazar jets is by imaging the radio sources
using the technique of Very Long Baseline Interferometry (VLBI). However, many HBLs are fainter
at radio wavelengths (typically tens of milli-janskys --- see Table~1) than the more
powerful quasars and BL Lac objects, and are not included in
VLBI monitoring programs such as MOJAVE \cite{lis16} and TANAMI \cite{ojh10}.
Properties that can be measured from VLBI
images --- the apparent jet speed, radio core brightness
temperature, core dominance, and jet--to--counter-jet brightness ratio ---
provide information on fundamental properties of the jet, such as
the bulk Lorentz factor and viewing angle.

Yet despite the high Doppler factors inferred from TeV observations, our previous VLBA
observations have established that TeV sources have only modest
brightness temperatures, and jet component motions that are sub-luminal or
only slightly superluminal
\cite{pin99,edw02,gir04,pin04,pin05,pin08,pin09,tie12,pin14}.
This has been called the``doppler crisis'' \cite{tav06} or
``bulk Lorentz factor crisis'' \cite{hen06}.

\begin{table}[H]
  \caption{Current status of our VLBA monitoring program. TeV source
    names are those used by TeVCat, redshifts are those given in \cite{pin14}, where an
    asterisk denotes a tentative value or limit. The NVSS flux density is
    measured at 1.4 GHz. The number of epochs refers to the number of
    VLBA images made in our monitoring program. References are to
    papers presenting these image.}  \centering
\begin{tabular}{lllrcc}
\toprule
\textbf{TeV source name} & \textbf{Redshift} & \textbf{NVSS counterpart} & \textbf{NVSS flux} & \textbf{Number} & \textbf{Refs}\\
\textbf{}                & \textbf{}         & \textbf{}                 & \textbf{density (mJy)}	& \textbf{of epochs} & \textbf{}\\
\midrule
SHBL J001355.9$-$185406       &  0.094     & NVSS J001356$-$185406 &    29.2 &   4 &   \cite{pin14}  \\ 
KUV 00311$-$1938              &  0.506 *   & NVSS J003334$-$192133 &    18.5 &   4 &   \cite{pin14}  \\ 
1ES 0033+595                  &  0.240 *   & NVSS J003552+595005   &   147.3 &   5 &   \cite{pin14}  \\ 
RGB J0136+391                 &  0.400 *    & NVSS J013632+390559   &    60.0 &   4 &   \cite{pin14}  \\ 
RGB J0152+017                 &  0.080     & NVSS J015239+014717   &    61.4 &   5 &   \cite{pin14}  \\ 
1ES 0229+200                  &  0.140     & NVSS J023248+201716   &    82.4 &   5 &   \cite{pin13,pin14}  \\ 
PKS 0301$-$243                &  0.266     & NVSS J030326$-$240710 &   700.2 &   2 &   ---   \\ 
IC 310                        &  0.019     & NVSS J031642+411928   &   168.1 &   2 &   ---   \\ 
RBS 0413                      &  0.190     & NVSS J031951+184536   &    20.9 &   5 &   \cite{pin13,pin14}  \\ 
1ES 0347$-$121                &  0.188     & NVSS J034922$-$115914 &    23.9 &   5 &   \cite{pin13,pin14}  \\ 
1ES 0414+009                  &  0.287     & NVSS J041652+010526   &   119.6 &   5 &   \cite{pin13,pin14}  \\ 
1ES 0502+675                  &  0.314     & NVSS J050755+673724   &    25.4 &   5 &   \cite{pin13,pin14}  \\ 
PKS 0548$-$322                &  0.069     & NVSS J055040$-$321620 &   343.7 &   5 &   \cite{pin13,pin14}  \\ 
RX J0648.7+1516               &  0.179     & NVSS J064847+151625   &    64.2 &   4 &   \cite{pin14}  \\ 
1ES 0647+250                  &  0.450     & NVSS J065046+250259   &    96.2 &   5 &   \cite{pin14}  \\ 
RGB J0710+591                 &  0.125     & NVSS J071030+590817   &   158.4 &   5 &   \cite{pin13,pin14}  \\ 
1ES 0806+524                  &  0.138     & NVSS J080949+521858   &   182.4 &   3 &   \cite{pin13}  \\ 
RBS 0723                      &  0.198     & NVSS J084712+113350   &    32.8 &   1 &   ---   \\ 
1RXS J101015.9$-$311909       &  0.143     & NVSS J101015$-$311906 &    73.5 &   4 &   \cite{pin14}  \\ 
1ES 1011+496                  &  0.212     & NVSS J101504+492601   &   377.7 &   3 &   \cite{pin13}  \\ 
1ES 1101$-$232                &  0.186     & NVSS J110337$-$232924 &   120.3 &   5 &   \cite{tie12}  \\ 
Markarian 421                 &  0.031     & NVSS J110427+381232   &   767.4 &  17 &   \cite{pin99,pin05,pin10}  \\ 
Markarian 180                 &  0.045     & NVSS J113626+700925   &   327.1 &   7 &   \cite{tie12}  \\ 
RX J1136.5+6737               &  0.134     & NVSS J113629+673706   &    45.3 &   1 &   ---   \\ 
1ES 1215+303                  &  0.130     & NVSS J121752+300700   &   571.6 &   2 &   ---   \\ 
1ES 1218+304                  &  0.184     & NVSS J122121+301036   &    71.0 &   5 &   \cite{tie12}  \\ 
MS 1221.8+2452                &  0.218     & NVSS J122424+243623   &    24.5 &   4 &   \cite{pin14}  \\ 
H 1426+428                    &  0.129     & NVSS J142832+424022   &    58.0 &   5 &   \cite{pin08,pin10}  \\ 
1ES 1440+122                  &  0.163     & NVSS J144248+120040   &    68.8 &   4 &   \cite{pin14}  \\ 
PG 1553+113                   &  0.500 *   & NVSS J155543+111124   &   312.0 &   7 &   \cite{tie12}  \\ 
Markarian 501                 &  0.034     & NVSS J165352+394536   &  1558.0 &   9 &   \cite{edw02,pin09,pin10}  \\ 
H 1722+119                    &  0.170 *   & NVSS J172504+115215   &   120.4 &   4 &   \cite{pin14}  \\ 
1ES 1727+502                  &  0.055     & NVSS J172818+501311   &   200.7 &   2 &   ---   \\ 
1ES 1741+196                  &  0.083     & NVSS J174357+193508   &   301.2 &   4 &   \cite{pin14}  \\ 
1ES 1959+650                  &  0.047     & NVSS J195959+650854   &   249.6 &  13 &   \cite{pin04,pin05,pin10}  \\ 
PKS 2155$-$304                &  0.116     & NVSS J215852$-$301330 &   489.3 &   7 &   \cite{pin04,pin05,pin10}  \\ 
B3 2247+381                   &  0.119     & NVSS J225005+382437   &   103.4 &   4 &   \cite{pin14}  \\ 
1ES 2344+514                  &  0.044     & NVSS J234705+514217   &   250.4 &   8 &   \cite{pin04,pin10}  \\ 
H 2356$-$309                  &  0.165     & NVSS J235907$-$303740 &    62.1 &   5 &   \cite{tie12}  \\ 
\bottomrule
\end{tabular}
\end{table}

Of the 46 HBLs detected at TeV energies:
\begin{itemize}[leftmargin=*,labelsep=4mm]
\item 11 have jet kinematics published previously by us. (Some of these are also in MOJAVE.) 
\item 7 have, or soon will have, speeds determined by the MOJAVE program \cite{lis16}.
\item 20 are included in the current phase of our program, with
first epoch VLBA results for all 20 published \cite{pin14}.
We present here the kinematic results for the first portion of these sources. 
\item 4 are too far south to be studied with the VLBA, with several of these
  are part of the TANAMI monitoring program \cite{ojh10}.
\item 4 are recent detections which are yet to be monitored with VLBI. 
  \end{itemize}

The 39 TeV sources for which we have at least one VLBA image are
listed in Table 1, together with their redshift, association in the
NRAO VLA Sky Survey (NVSS) catalog \cite{con98}, 1.4 GHz flux density,
and details of our VLBA observations. Images and data available at the
project website:
www2.whittier.edu/facultypages/gpiner/research/archive/archive.html .

\begin{figure}[H]
\centering
\includegraphics[width=5.8cm]{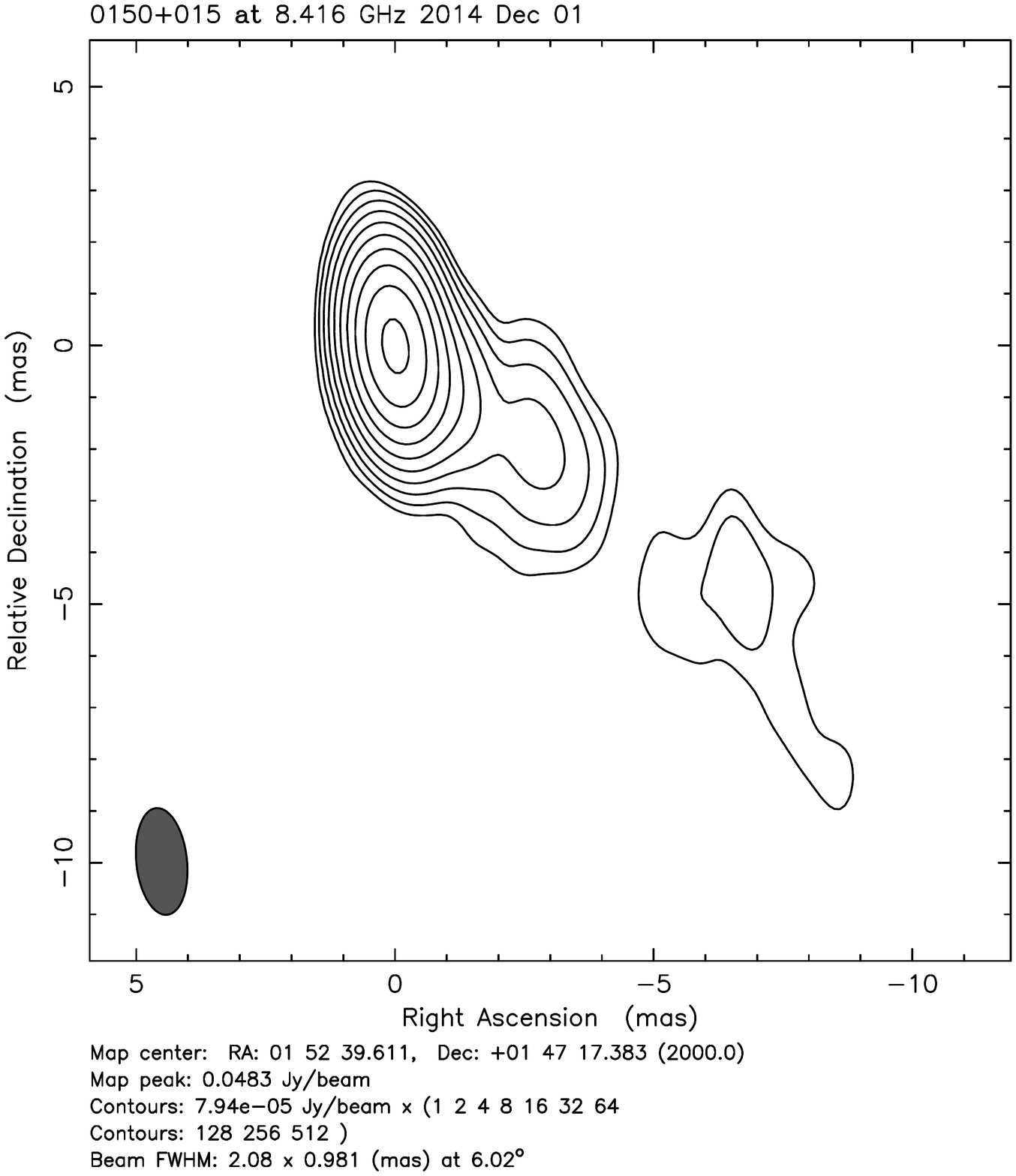} ~ ~ \includegraphics[width=5.8cm]{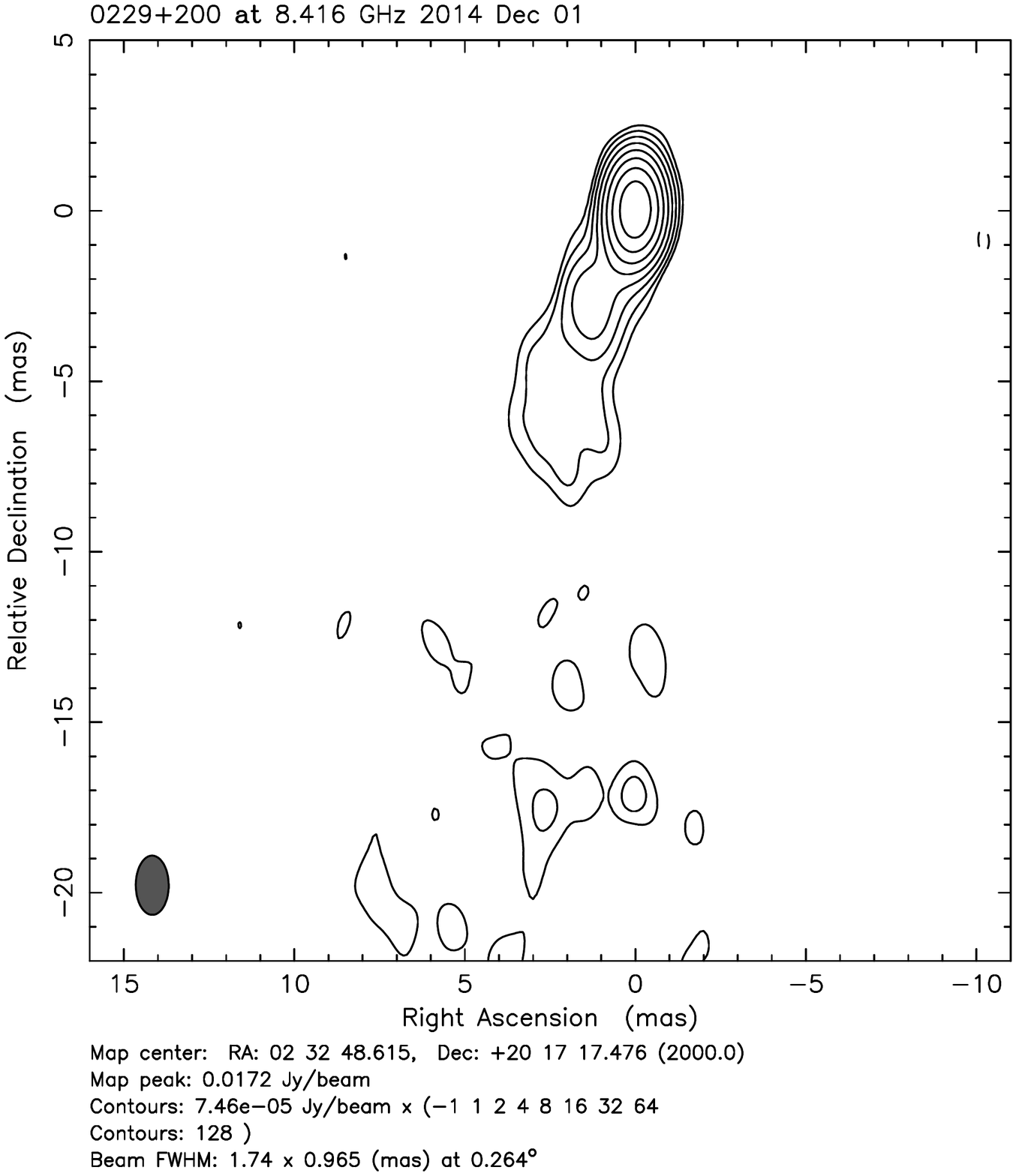}
\vspace*{5mm}
\includegraphics[width=5.8cm]{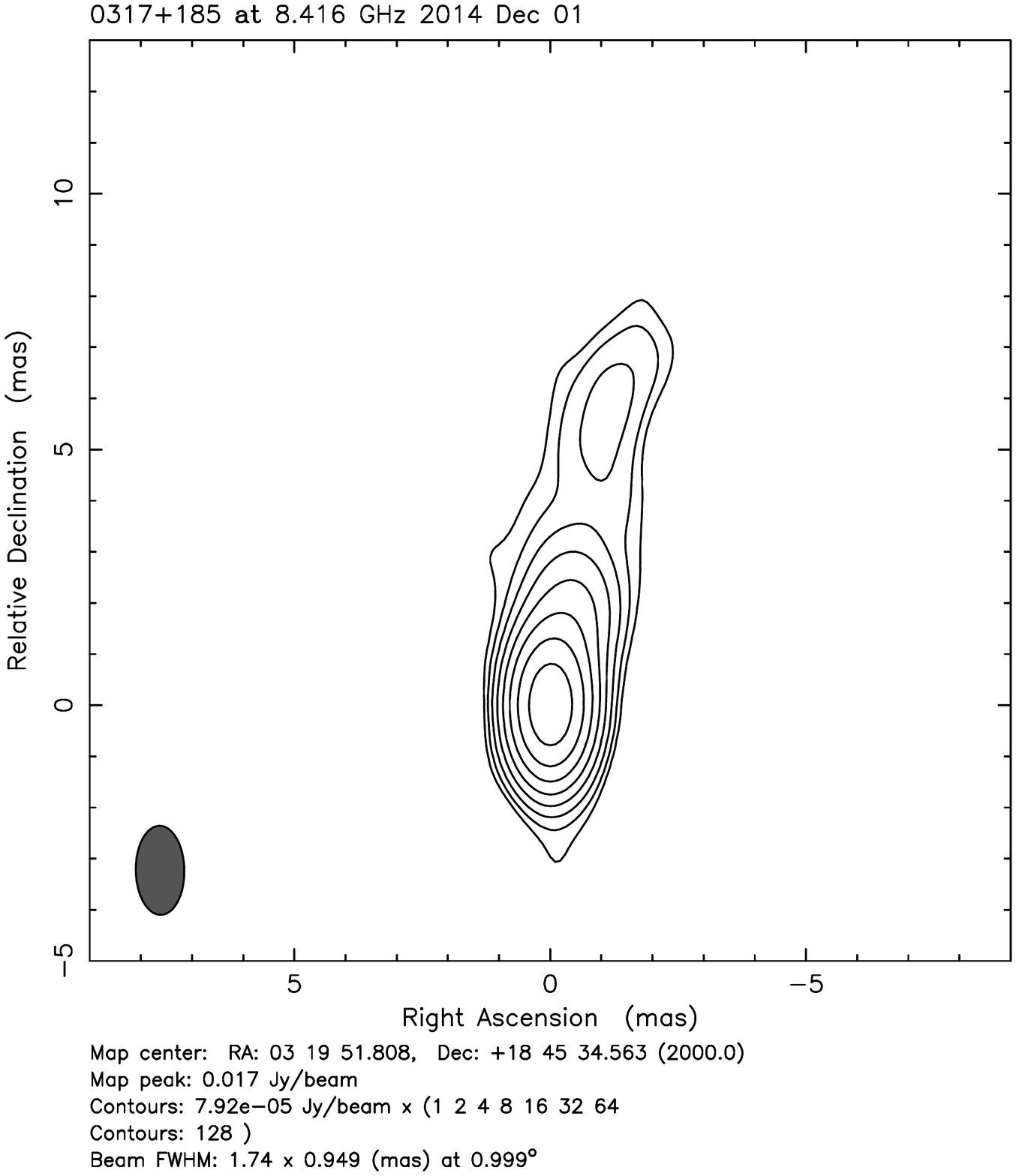} ~ ~ \includegraphics[width=5.8cm]{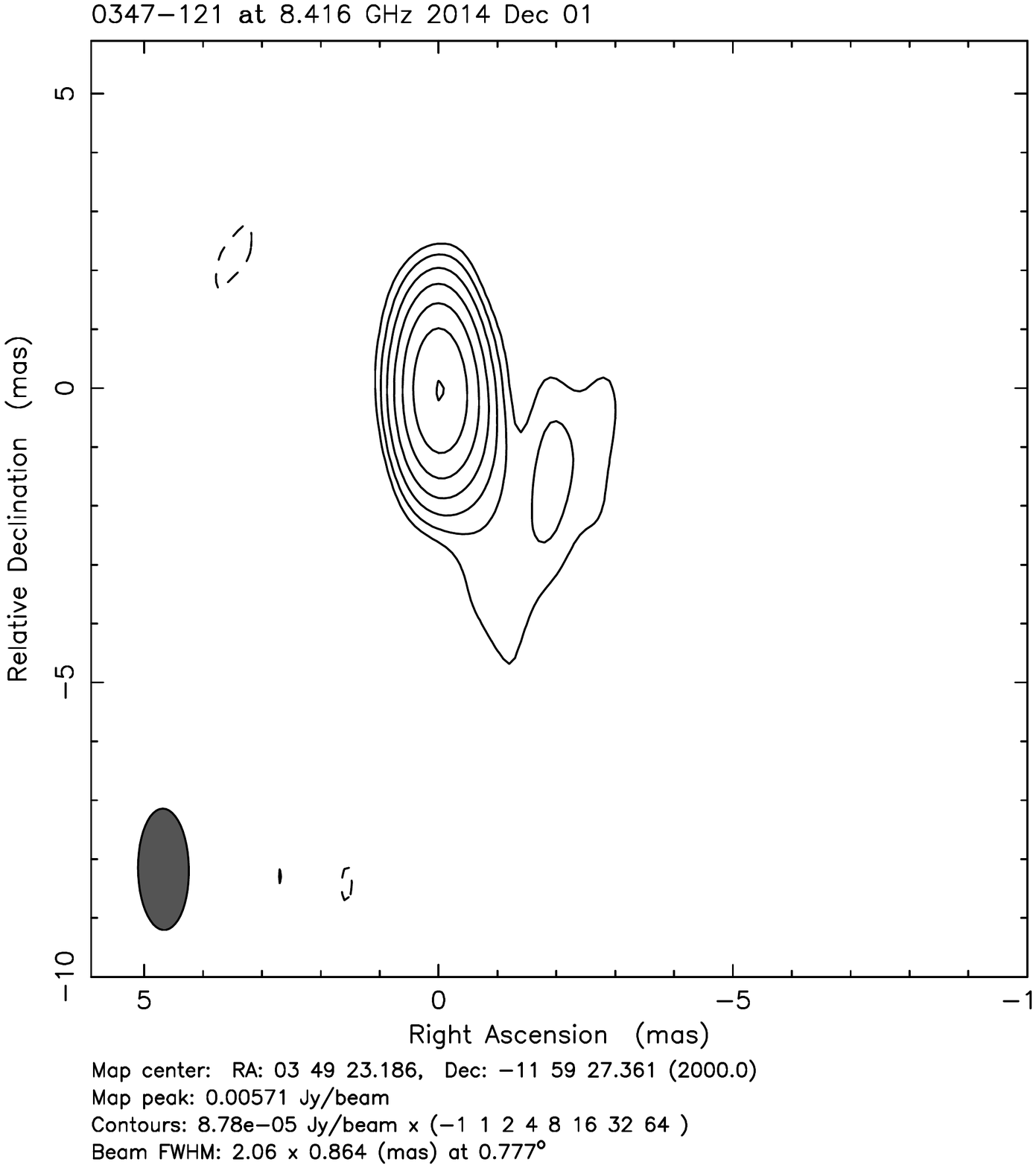}
\caption{VLBA images at 8\,GHz of four Tev blazars in out VLBA monitoring program:
(\textbf{a}) RGB J0152+017,  (\textbf{b}) 1ES 0229+200,   (\textbf{c}) RBS 0413 (0317+185),   (\textbf{d}) 1ES 0347$-$121.
  }
\end{figure}

TeV photons are attenuated by the infrared background \cite{aha06a}
and, as is apparent in Table~1, the majority of extragalactic TeV
sources are at relatively low redshift ($z <$ 0.2). Conversely,
studies of TeV gamma-ray spectra have offered a means of constraining
the infrared background \cite{aha06a,ste16}.

\section{Results}

Images of four of the sources currently being monitored are shown in
Figure~1.  All show parsec-scale morphologies typical of this class: a
compact core (which hosts the supermassive black hole powering the
source), and a weaker, one-sided jet that transitions to a
decollimated structure with larger opening angle at a few tens of
milli-arcseconds from the core (see image of 0229+200).  Multi-epoch
studies of these jets over the course of several years allow the
apparent speeds of the jet components to be determined.

Our previous VLBA studies indicated the absence of rapidly moving
features in the jets of TeV HBLs; jet components were either nearly
stationary or slowly moving ($< \sim1c$) \cite{tie12}.  With the
addition of multi-epoch data from eight previously unpublished
sources, including the four sources in Figure~1, the revised
distribution of apparent jet speeds is shown in Figure~2.  This Figure
incorporates the results of our previously published data and jet
speeds for four sources that have been monitored as part of the
MOJAVE project \cite{lis16}.

With the addition of new data, the tail of the distribution now
extends to mildy superluminal apparent speeds for the first
time; however, the majority of the TeV HBLs have peak apparent speeds
of only about 1$c$.
Combining these slow apparent speeds
with the high Doppler factors
($\delta$)
implied by the TeV data to
solve for the Lorentz factor
($\Gamma$)
and viewing angle ($\theta$) results in
unreasonbly small viewing angles ($\theta <<$ 1$^\circ$).
This would imply tiny jet opening angles, enormous
linear sizes, and huge numbers of parent objects,
and indicates that the
combination of both high Doppler factor and slow apparent speed in the
same jet region is unphysical.
If more realistic viewing
angles of a few degrees are assumed, then the observed apparent speeds imply
more modest Lorentz and Doppler factors for the radio jet.
The lack of detection of counterjets in the
VLBI images for any TeV HBL
\cite{gir08,pin14}
requires that the Doppler factor
cannot be arbitrarily low, and values of $\delta$ and $\Gamma$ of a few are most
consistent with the combined VLBI data.

\section{Discussion}

A variety of mechanisms have been proposed to try and
reconcile the Doppler crisis \cite{geo03,gop04,ghi05,gop06,gop07,mar14}.
The most natural explanation is for a range of Doppler
factors to coexist in the same jet on parsec scales through jet
stratification. One example is a jet that decelerates
along its length \cite{geo03}.
In such a jet, the fast inner part sees
blueshifted photons from the slower outer part, reducing the high
Lorentz factor required in the fast portion. This is a general feature
of models with velocity structures; radiative interaction among the
different regions allows the SED to be reproduced without the
extremely high Lorentz factors and Doppler factors characterizing
single-zone models. Another alternative is a transverse velocity structure 
with a fast central spine and a slower outer sheath.
Radiative interaction between the spine and sheath naturally
decelerates the spine, producing both radial and transverse velocity
structures in the same jet \cite{ghi05}.

\begin{figure}[H]
\centering
\includegraphics[angle=180,width=7cm]{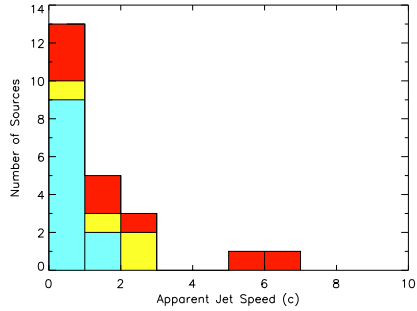}
\caption{A histogram of the peak jet speeds in TeV blazars.
Blue denotes sources from our previously published data, red are from this work, and yellow are from MOJAVE.
The highest apparent speed are observed in 
RBS\,0413, (6.0$\pm$1.2)$c$, and RGB\,J0710+591, (5.8$\pm$1.5)$c$.
}
\end{figure}

If such spine--sheath jets are present in TeV HBLs, then the outer layer
is expected to dominate the radio emission due to its SED shape, even
with a lower Doppler factor than the spine
\cite{ghi05}.
An observational
signature of this would be a limb-brightened transverse
profile for the jet in VLBI images.
There is evidence for limb-brightening close to the core
in a number of HBLs, e.g., Mkn 501 \cite{pin05},
1ES~0502+675 \cite{pin14}, and H~1722+119 \cite{pin14}.

Other possible jet velocity structures have also
been proposed, including
multiple blobs \cite{tav11}
fast moving “needles” within the main jet
\cite{ghi08}, ``minijets'' powered by magnetic
reconnection events \cite{gio13},
and turbulent subregions within the jet \cite{mar14}.


\section{Conclusions}

Our on-going VLBA monitoring of the growing number of
TeV gamma-ray emitting HBLs has revealed that
the distribution of peak apparent jet speeds in
these sources extends to moderate superluminal speeds,
$\sim$6$c$, but the majority display
subluminal speeds, in contrast with the distribution for
other classes of active galactic nuclei \cite{lis16}.
It has recently been proposed that jet kinematics may
offer a better classification for blazars than the
SED peak frequency \cite{her16}, with HBLs tending to
display quasi-stationary knots arising from recollimation shocks.

A possible physical explanation for this is based
on TeV blazars 
having intrinsically weak jets that interact with the external medium forming a slow
surrounding layer. Radiative interaction between the spine and the sheath
decelerates the spine, and eventually disrupts the jet. Such jets
are prominent in TeV-selected samples because selection 
favors rare high-synchrotron peak sources, which are drawn
from the low end of the luminosity function where the source density
is largest \cite{gio12}.

\vspace{6pt} 


\acknowledgments{
The National Radio Astronomy Observatory is a facility of the National Science Foundation operated under cooperative agreement by Associated Universities, Inc.
This research has made use the TeVCat online source catalog (http://tevcat.uchicago.edu).
This research has made use of NASA's Astrophysics Data System.
This research has made use of the NASA/IPAC Extragalactic Database (NED) which is operated by the Jet Propulsion Laboratory, California Institute of Technology, under contract with the National Aeronautics and Space Administration.}

\bibliographystyle{mdpi}



\bibliography{mybib}


\end{document}